\documentclass[12pt]{article}
\usepackage{amsmath}
\usepackage{epsf}

\newcounter{figset}
\newcounter{subfig}

\newenvironment{subfigure}{
           }{
           }
\input{tcilatex}

\begin{document}

\title{Analysis and classification of nonlinear dispersive evolution
equations in the potential representation}
\author{ U.A.\ Eichmann,  J.P.\ Draayer \\
{\normalsize {Department of Physics and Astronomy, Louisiana State
University,} }\\
{\normalsize {Baton Rouge, LA 70803-4001} } \and  and A. Ludu \\
Dept. Chemistry and Physics,\\
Northwestern State University, Natchitoches, LA 71497}
\date{}
\maketitle

\begin{abstract}
\noindent A potential representation for the subset of traveling solutions
of nonlinear dispersive evolution equations is introduced. The procedure
involves a reduction of a third order partial differential equation to a
first order ordinary differential equation. In this representation it can be
shown that solitons and solutions with compact support only exist in systems
with linear or quadratic dispersion, respectively. In particular, this
article deals with so the called $K(n,m)$ equations. It is shown that these
equations can be classified according to a simple point transformation. As a
result, all equations that allow for soliton solutions join the same
equivalence class with the Korteweg-deVries equation being its
representative.
\end{abstract}

\vskip2cm

\section{Introduction}

The description of a physical system under extreme conditions, e.g. large
amplitude excitations, requires more than a linear theory. However, since
dissipation is normally only effective over large time scales, it can be
neglected, at least in a first approximation, leaving a nonlinear dispersive
partial differential equation for describing the behavior of such a system.
In particular, one typically has to deal with third-order as well as
first-order spatial derivatives of a wave function $u$ or powers of $u$,
i.e.\ $(u^m)_{xxx}$, which describe (nonlinear) dispersion, and $(u^n)_x$,
which describes nonlinear convection, plus a dynamical term, namely, the
time derivative of the wave function, $u_t$.

A special class of nonlinear dispersive evolution equations in $1+1$
dimensions are the so-called $K(n,m)$ equations, $(u^m)_{xxx} - A (u^n)_x +
u_t = 0$ \cite{Rosenau-Hyman}. The most prominent examples of these are the
Korteweg-deVries (KdV) equation, $K(2,1)$, and the modified KdV equation, $%
K(3,1)$. These forms are of particular interest because they can be used to
describe the motion of stable and localized solitary waves (solitons) that
are observed in a variety of physical systems. These forms are also of
particular import for various specialized applications, such as data
transfer in fiber \cite{Agrawal} or as an analytical tool for an explanation
of cluster radioactivity \cite{Ludu-cluster} or nuclear fission \cite%
{Draayer-fission}.

The considerations in this article will be carried out for the specific case
of traveling wave solutions. It will be shown that this restriction leads to
a special representation of nonlinear dispersive evolution equations called
the potential representation. (This concept has been used previously, for
example, in \cite{Rosenau-potrep,Rosenau-compact}). This representation
resembles an energy conservation law with a nonrelativistic kinetic energy
term as well as a potential energy. The potential representation, being a
first order ordinary differential equation, constitutes an enormous
simplification of the original problem. Gross properties of the solutions
can be read directly from the potential function without the actual need to
solve a differential equation. The conditions for solitary waves and
solitons can thus be easily stated qualitatively. Moreover, the
investigation of the potential picture directly reveals that solitons can
only emerge in systems with linear dispersion. Compactons, i.e. solitary
waves with compact support, only exist in systems with quadratic dispersion.
These results are discussed in more detail for the specific cases of systems
that are modeled by $K(n,m)$ equations.

By restricting our consideration to a particular subset of solutions, a $%
K(n,m)$ equation can be transformed into another $K(N,M)$ equation by means
of a simple point transformation. This point transformation defines an
equivalence relation between the various $K(n,m)$ equations and in so doing
divides these equations into equivalence classes of connected equations.

\section{Potential Representation}

\subsection{Reformulation of the nonlinear evolution equation}

We consider a general nonlinear dispersive evolution equation of an
autonomous one-dimensional non-dissipative dynamical system 
\begin{equation}  \label{gen-eq}
\left(u^m\right)_{xxx} = V(u) u_x - u_t\quad ,
\end{equation}
where $V(u)$ is an arbitrary integrable and continuous function of $u$. The
subscripts denote partial differentiation with respect to the index. We
focus on traveling solutions with the space-time dependence $u(x,t)= u(x-vt)
= u(\xi)$, with $v$ denoting the speed, so that the partial differential
equation (\ref{gen-eq}) is reduced to the ordinary differential equation 
\begin{equation}  \label{gen-ode}
\left(u^m\right)_{\xi\xi\xi} =V(u) u_\xi + v u_\xi = \mathcal{V}(u)
u_\xi\quad ,
\end{equation}
Equation (\ref{gen-ode}) can be integrated once 
\begin{equation}  \label{zwschritt1}
\left(u^m\right)_{\xi\xi} = \int_0^u \mathrm{d}t \mathcal{V}(t) - C_1 \quad ,
\end{equation}
and by using an integrating factor $\left(u^m\right)_\xi$ (see Appendix \ref%
{ifappend}) eq.\ (\ref{zwschritt1}) turns into 
\begin{equation}
\frac{m^2}{2}\left[ u^{2m -2} \left(u_\xi\right)^2 \right]_\xi = \left[%
\int_0^{u(\xi)} \mathrm{d}t \; m\;t^{m-1} \int _0 ^t \mathrm{d}s \mathcal{V}%
(s)\right]_\xi - C_1 \left( u^m\right) _\xi\quad ,
\end{equation}
allowing a further integration and leading finally to 
\begin{equation}  \label{potrep}
\left(u_\xi\right)^2 = -\mathcal{F}(u)\quad ,
\end{equation}
where $\mathcal{F}(u)$ is given by 
\begin{equation}  \label{F-V}
\mathcal{F}(u) = C_1 u^{2-m} + C_2 u^{2-2m} - \frac{2}{m^2} u^{3-m} \int_0^1 
\mathrm{d}x (1-x^m) \mathcal{V}(ux)\quad .
\end{equation}
In the following, eq.\ (\ref{potrep}) is referred to as the \textit{%
potential representation} of eq.\ (\ref{gen-eq}). This notation is inspired
by the fact that if $\xi$ and $u$ are identified with \textit{time} and 
\textit{space}, respectively, the l.h.s.\ of eq.\ (\ref{potrep}) may be
associated with a nonrelativistic `kinetic' energy, and, accordingly, the
r.h.s.\ with the negative value of a `potential' energy $\mathcal{F}(u)$. $%
u_\xi$ is then the `velocity' of a particle moving along the $u$-axis. Only
nonlinear evolution equations of the form (\ref{gen-eq}) can be reduced to
this potential form. This integrability property is related to the existence
of an `energy' conservation law. The evolution of $u$ proceeds on the
zero-energy hypersurface in the phase space belonging to $\mathcal{F}(u)$. $%
\mathcal{V}(u)$ can now be expressed as 
\begin{equation}
\mathcal{V}(u)=- m(m-1)(m-2) u^{m-3} \mathcal{F}(u) - \frac32 m(m-1) 
\mathcal{F}^\prime (u) - \frac 12 m u^{m-1} \mathcal{F}^{\prime\prime}(u)%
\quad ,
\end{equation}
where the prime indicates differentiation with respect to the argument.

The problem of integrating eq.\ (\ref{gen-eq}) is thus reduced to a
quadrature that can be solved by separation of variables yielding $\xi$ as a
function of $u$. One is now left with the challenging task of inverting this
function, which in many cases may not be possible.

For later purposes we briefly mention that the potential function for
evolution equations of the form 
\begin{equation}
\left((u+\gamma)^m\right)_{xxx} = \mathcal{V}(u) u_x
\end{equation}
with an arbitrary constant $\gamma$ is given by 
\begin{equation}  \label{pfshift}
\mathcal{F}(u) = C_1 (u+\gamma)^{2-m} + C_2 (u+\gamma)^{2-2m} - \frac{2}{m^2}
(u+\gamma)^{3-m} \int_0^1 \mathrm{d}x (1-x^m) \mathcal{V}(ux-(1-x)\gamma)%
\quad .
\end{equation}

\subsection{Analysis of the potential representation}

\label{potanalys} Casting the original nonlinear dispersive evolution
equation in the potential representation associates to the wave function $%
u(\xi)$ a space-time (i.e.\ $u-\xi$) trajectory of a particle moving in the
potential $\mathcal{F}(u)$ with zero total energy. Different types of
solutions of the original nonlinear evolution equation can be attributed to
different kinds of trajectories in this phase space. For example, closed
trajectories in bounded regions of the phase space correspond to periodic
solutions, whereas solitary wave solutions are represented in phase space by
separatrix trajectories \cite{Arnold}. In the following, the properties and
conditions of solitary waves and of solitary waves with compact supportare
discussed in more detail. However, the so-called \textit{kinks}, solitary
waves which are represented in phase space by separatrix trajectories with
two cusps, are not considered.

\subsubsection*{Solitons}

In the potential representation, necessary and sufficient conditions for
solitary waves read: 
\begin{eqnarray}  \label{soli-cond}
\mathcal{F}(a) = \mathcal{F}^\prime (a) = 0,\quad \mathcal{F}%
^{\prime\prime}(a)\leq 0 ,  \notag \\
\mathcal{F}(b) = 0, \quad \mathcal{F}^\prime (b) \neq 0, \quad\quad a<b\quad
.
\end{eqnarray}
That is, the potential function must have at least a two-fold zero at a
point $u=a$ with negative or vanishing curvature and must have a zero at $%
b>a $ with nonvanishing slope. In addition, $\mathcal{F}(u)$ must not have
singularities in the interval $[a,b]$, and therefore $\mathcal{F}^\prime (b)$
is greater than $0$. The reasons underlying (\ref{soli-cond}) are as
follows: A particle at $u=a$ is at rest (the potential energy is equal to
the total energy) and does not experience any force (the gradient of the
potential is zero). It takes the particle infinitely long to leave this
point. It moves in a positive $u$ direction, is reflected at $u=b$, moves
back in a negative $u$ direction and reaches $a$ again after another
infinite time span, i.e.\ $\lim_{|\xi|\to \infty} u(\xi) \to a$, $u(0)=b$.
The simple zero of the potential function, $b$, thus corresponds to the
amplitude of the solitary wave.

Localized solitary waves (solitons) require $a = 0$. For dark solitary
waves, i.e.\ solitary waves with negative amplitude, $\mathcal{F}(u)$ has to
fulfill (\ref{soli-cond}) with $b<a$ (the expression \textit{dark soliton}
has been introduced in the context of nonlinear optical pulse propagation %
\cite{hasegawa}). The above conditions for solitons lead to $C_1 = C_2 = 0$
in (\ref{F-V}). One also finds in general as a necessary premise for soliton
solutions resulting from equations of the form (\ref{gen-eq}) that $m=1$
which is a consequence of the dynamical term $u_t$.

Focusing on $K(n,m)$-type equations with 
\begin{equation}
\mathcal{V}(u) = n A u^{n-1} +v
\end{equation}
one finds for the potential function 
\begin{equation}  \label{potknm}
\mathcal{F}(u) = -\frac{2A}{m(m+n)}u^{n-m+2} - \frac{2v}{m(m+1)}u^{3-m} +
C_1 u^{2-m} + C_2 u^{2-2m}\quad .
\end{equation}
The application of the conditions dicussed above for solitons explicitely to
this potential function yield besides $C_1=C_2=0$ 
\begin{equation}
\begin{array}{lcl}
v & > & 0,\;\; A<0 \\ 
b & = & \displaystyle \left(\frac {v(n+1)}{2 |A|}\right)^{\frac1{n-1}}%
\end{array}%
\end{equation}
That is, the soliton moves in a positive $x$ direction ($v>0$) and the
parameter $A$ has to be smaller than zero. The amplitude of the soliton is
proportional to the $(n-1)$-th root of the velocity. For dark solitons,
i.e.\ antisolitons, one finds instead $A>0$ for $n$ even and $A<0$.

Finally one can deduce basic properties of the width $L$ of the soliton. The
width is calculated at a certain height of the soliton, e.g.\ at half the
maximum height 
\begin{equation}
L= 2\int_{b/2}^b \frac{\mathrm{d}u}{\sqrt{-\mathcal{F}(u)}} = 2\int_{b/2}^b 
\frac{\mathrm{d}u}{\sqrt{\frac{2 A}{n+1}u^{n+1} + v u^2}}= \frac 2{\sqrt{v}}
\int_{1/2}^1 \frac{\mathrm{d}u}{\sqrt{u^{n+1} + u^2}}
\end{equation}
The last step follows with $2 A/(v(n+1)) = b^{1-n}$. One thus finds that 
\begin{equation}
L\sim \frac 1{\sqrt{v}}= \sqrt{\frac{n+1}{2|A|b^{n-1}}}
\end{equation}
for any $n$.

The soliton solutions of $K(n,1)$ equations have the form (see section \ref%
{kdv-ec}) 
\begin{equation}
u(\xi) = (\pm)^n \left(\frac{(n+1) v}{2 A}\right) ^{\frac{1}{n-1}} \frac{1}{%
\cosh^\frac{2}{n-1}\left(\frac{(n-1)\sqrt{v}}{2}\xi\right)}
\end{equation}

\subsubsection*{Compact Support Solutions}

In this section we will discuss localized solitary waves with compact
support, i.e. 
\begin{equation}
u(\xi)\left\{ 
\begin{array}{rcl}
& \neq & 0 \;\mathrm{if}\; \xi \in [\xi_1,\xi_2] \\ 
& = & 0\; \mathrm{else}%
\end{array}%
\right.
\end{equation}
which here shall be called \textit{compact support solutions} (CSS). These
solutions were introduced as \textit{compactons} in \cite{Rosenau-Hyman}. $%
u(\xi)$ being a solution of (\ref{gen-eq}) requires $u^m(\xi)$ to be of
class $C_3$ and $u(\xi)$ to be at least of class $C_1$ at the boundary of
the compact $\xi$ interval. Since unlike the soliton the CSS does not
approach zero asymptotically but reaches it in a finite time $\xi$, the
gradient of the potential function must never vanish in the interval $0\leq
u\leq b$ ($b$ denotes the amplitude of the CSS), or, equivalently, $%
u_{\xi\xi}\neq 0$ for all $\xi \in [\xi_1,\xi_2]$. This leads to 
\begin{eqnarray}  \label{comp-cond}
\mathcal{F}(0) &=&0, \; \mathcal{F}^\prime(0) <0  \notag \\
\mathcal{F}(b) &=&0, \; \mathcal{F}^\prime (b)>0
\end{eqnarray}
From the requirements for the wavefunction and its derivatives at the
boundary of the finite $\xi$ interval one obtains that either $m=2$ or $%
m\geq 3$. However, with (\ref{comp-cond}) one finds $m=2$ and $C_1=C_2=0$.
The conditions for compactons can finally be stated as follows 
\begin{equation}  \label{compheight}
\begin{array}{l}
m=2 \\ 
v>0,\; A<0 \\ 
\displaystyle b=\left(\frac{v (n+2)}{3 |A|}\right)^{\frac{1}{n-1}}%
\end{array}%
\end{equation}
Similar to the soliton case, the amplitude is proportional to the $(n-1)$-th
root of the velocity. CSS with negative amplitude require both $v<0$, i.e.\
they are moving in negative $x$-direction, and either $A>0$ for $n$ odd or $%
A<0$.

In the same way as for solitons one finds for the widths of a CSS the
following relation 
\begin{equation}  \label{compwidth}
L \sim \sqrt{\frac{b}{v}}=\sqrt{\frac13}\left(\frac v3\right)^{\frac{ 2-n}{%
2(n-1)}}\left(\frac {|A|}{(n+2)}\right)^{\frac{ -1}{2(n-1)}}
\end{equation}
For $n=2$ one thus finds the known result, that the width of the CSS is
independent of its speed or its height \cite{Rosenau-Hyman}.

As examples for CSS we give the solutions of the $K(2,2)$ equation 
\begin{equation}  \label{k22solu}
u (\xi) =\left\{%
\begin{array}{ll}
\displaystyle \frac{4v }{3 A } \cos^2 \left( \frac{\sqrt{A }}{4}\xi\right) & 
\mathrm{for} \quad |\sqrt{A }\xi/4|\leq \pi \\ 
0 & \mathrm{otherwise}%
\end{array}
\right.
\end{equation}
and of the $K(3,2)$ equation 
\begin{equation}
u(\xi) =\left\{%
\begin{array}{ll}
\displaystyle \sqrt{\frac{5v}{3|A|}}\mathrm{sn}^2\left(\left.\left(\frac{v
|A|}{240}\right)^{\frac14}\xi\right|-1\right) & \mathrm{for}\quad %
\displaystyle 0\leq \xi\leq \left(\frac{240}{v|A|}\right)^{\frac14} \frac2{%
\sqrt{2}}K\left(\frac12\right) \\ 
0 & \mathrm{otherwise}%
\end{array}
\right.
\end{equation}
both clearly showing properties (\ref{compheight}) and (\ref{compwidth})
found from the general analysis of the potential function. In the latter
solution, $\mathrm{sn}(x|m)$ denotes the Jacobian elliptic function $\mathrm{%
sn}(x|m) = \sin (\mathrm{am}\, u)$ with the Jacobi amplitude $\mathrm{am}
\,u $ and $u=u(x, m) = \int_0^x \mathrm{d}\theta (1-m\sin^2\theta)^{-1/2}$. $%
K(m) $ denotes the quarter period $u(\pi/2,m)$ \cite{Abramowitz}.

Particularly for the $K(2,2)$ case, we want to mention that a solution (\ref%
{k22solu}) added to a constant $\delta$ is again a solution of a $K(2,2)$
equation. If $u(\xi)$ solves the $K(2,2)$ equation $(u^2)_{\xi\xi\xi} = A
(u^2)_\xi + v u_\xi$ with the potential function (see eq.\ (\ref{potknm})) 
\begin{equation*}
\mathcal{F}(u) = -\frac{A}{4} u^2 - \frac{v}{3} u
\end{equation*}
then $U(\xi)=u(\xi) +\delta$ obeys, according to (\ref{pfshift}), the
following potential representation 
\begin{equation}  \label{potrepk22shift}
(U_\xi)^2 = \frac{A}{4} U^2 + (\frac v3 + \frac{A}{2}\delta) U - \frac{A}{4}%
\delta^2 - \frac v3 \delta
\end{equation}
for the transformed equation $((U-\delta)^2)_{\xi\xi\xi} = A (U^2)_\xi +
(v-2A\delta)U_\xi$ with $C_1=C_2=0$. Eq.\ (\ref{potrepk22shift}) may,
however, as well be interpreted as the potential representation of a $K(2,2)$
equation with $C_1 = - A\delta^2/4 - v \delta/3$. The solution $U(\xi)$
moves with the velocity $V = v + 3/2 |A| \delta = 3/4|A| (b + 2\delta)$, $v$
being the velocity of $u(\xi)$ \cite{Ludu-OSA}. \\[0.1cm]

The above results show that in nonlinear dispersive, non-dissipative
dynamical systems soliton solutions may only occur if the dispersion is
linear ($m=1$) whereas compact support solutions require a quadratic
dispersion ($m=2$). This is summarized in Fig.\ 1 showing a chart of $K(n,m)$
equations.

\section{Classification of $K(n,m)$-equations by a point transformation}

\subsection{General considerations}

In this section it will be shown that $K(n,m)$-type equations can be
transformed into other $K(N,M)$-type equations with different arguments $N$
and $M$. The applied (point) transformation uniquely connects the elements
of certain sets of $K(n,m)$ equations. It constitutes an equivalence
relation between those connected equations and thus serves as a tool to
classify nonlinear dispersive evolution equations of the $K(n,m)$ type. The
benefit of this classification is clearly the fact that solutions to any
element (equation) of an equivalence class can be traced back to the
solution of the representative of this class. However, this transformation
requires to restrict the considerations to a certain subset of traveling
solutions which is defined by fixing the initially arbitrary integration
constants in the potential representation $C_1=C_2=0$. In the case of $%
K(n,1) $ or $K(n,2)$ equations this is the necessary condition for solitons
and compact support solutions.

Transformations of the potential representation, being a first order ODE,
are fully covered by the theory of point transformations \cite{Anderson}.
Here we choose the transformation $u(\xi) = \omega^{\pm q}(\xi)$ which
belongs to a certain class of point transformations that transport the
potential representation of a nonlinear dispersive differential equation of
type $K(n,m)$ into the potential representation of another $K(N,M)$
equation. The general derivation of admissible transformations is carried
out in Appendix \ref{ptderiv}. The transformation presented above has been
chosen for further considerations because it allows for an entirely
analytical treatment.

This simple point transformation suggests diagram (Fig.\ \ref{diag1}) in
which the $K(n,m)$ and the $K(N,M)$ equation are connected via the
respective potential representations. The subscript $0$ of the potential
function indicates the special choice $C_1=C_2=0$. Thus, if $u(\xi)$ is a
solution of the $K(n,m)$ equation, $\omega (\xi)$ is a solution of the $%
K(N,M)$ equation. \setcounter{figset}{\value{figure}} \setcounter{subfig}{2} %
\setcounter{figure}{0}

\begin{subfigure}
The integers $M$ and $N$ fulfill the relations 
\begin{eqnarray}  \label{trafo-}
\left. 
\begin{array}{lcl}
M & = & q(m-1) +1 \\ 
N & = & q(n-1)+1%
\end{array}%
\right\}&\quad\quad& \mathrm{for}\;\;u\to \omega^{+q} \\
\left. 
\begin{array}{lcl}
M & = & q(n-m) +1 \\ 
N & = & q(n-1)+1%
\end{array}%
\right\}&\quad\quad& \mathrm{for}\;\;u\to \omega^{-q}
\end{eqnarray}
For $M$, $N$, $m$ and $n$ being integers, $q$ can take on any rational
number $q=(N-1)/(n-1)$ but in fact, the diagram remains valid for any value
of $q$. Note, however, that although in this way the $K(n,m)$ equations may
be transformed into equations with arbitrary, infinitesimal nonlinearities $%
n,\;m =1+\varepsilon$, the resulting equations do not represent analytic
continuations of the linear cases and the respective solutions are not
smoothly connected.

The $K(n,m)$ equations with $m=(n+1)/2$ are, as a peculiarity of the
transformation, again transformed into equations with $M=(N+1)/2$ by both (%
\ref{trafo+}) and (\ref{trafo-}).
\end{subfigure}

\setcounter{figure}{\value{figset}} \addtocounter{figure}{1}

The equation $K(n,m)$ can be directly transformed into $K(N,M)$ by using the
potential representation of the $K(N,M)$ equation as a consistency relation.
The diagram (Fig.\ 2) can thus be closed (see Fig.\ 3). To see this one
calculates explicitely for the case (\ref{trafo+}) with $u=\omega ^{+q}$ 
\begin{eqnarray}
\left( u^{m}\right) _{\xi \xi \xi } &=&A\left( u^{n}\right) _{\xi }+vu_{\xi }
\notag  \label{knmu} \\
\left( \omega ^{mq}\right) _{\xi \xi \xi } &=&\left( \omega ^{q-1+M}\right)
_{\xi \xi \xi }=A\left( \omega ^{nq}\right) _{\xi }+v\left( \omega
^{q}\right) _{\xi }
\end{eqnarray}%
Expanding $(\omega ^{q-1+M})_{\xi \xi \xi }$ eq.\ (\ref{knmu}) can be
rearranged 
\begin{eqnarray}  \label{newknm}
\left( \omega ^{M}\right) _{\xi \xi \xi } &=&-M(q-1)\left( q+2M-4\right)
\omega ^{M-3}\left( \omega _{\xi }\right) ^{2}\omega _{\xi }-\frac{3}{2}%
M(q-1)\omega ^{M-2}\left[ \left( \omega _{\xi }\right) ^{2}\right] _{\xi } 
\notag  \label{trafodispers} \\
&&+\frac{AM}{\omega ^{q-1}mq}\left( \omega ^{nq}\right) _{\xi }+\frac{vM}{%
\omega ^{q-1}mq}\left( \omega ^{q}\right) _{\xi } \\
&\overset{!}{=}&A^{\prime }\left( \omega ^{N}\right) _{\xi }+v^{\prime
}\omega _{\xi }\quad .
\end{eqnarray}%
\TEXTsymbol{>}From the requirement that (\ref{trafodispers}) equals a $%
K(N,M) $ equation with coefficients $A^{\prime }$ and $v^{\prime }$ (\ref%
{newkmn}) we infer also the existence of a potential representation 
\begin{equation}
\left( \omega _{\xi }\right) ^{2}=-\mathcal{F}(\omega )=\frac{2A^{\prime }}{%
M(M+N)}\omega ^{N-M+2}+\frac{2v^{\prime }}{M(+1)}\omega ^{3-M}\;,\quad \left[
\left( \omega _{\xi }\right) ^{2}\right] _{\xi }=-\mathcal{F}^{\prime
}(\omega )\omega _{\xi }\;.
\end{equation}%
Here we have made the special choice $C_{1}=C_{2}=0$ which is a necessary
condition for the general case (see also appendix \ref{ptderiv}). Inserting
this condition for the squared derivative of the transformed wavefunction $%
\omega $ into (\ref{trafodispers}) one gets 
\begin{eqnarray}
\left( \omega ^{M}\right) _{\xi \xi \xi } &=&\frac{MA\;n}{m}\omega
^{N-1}\omega _{\xi }+\frac{Mv}{m}\omega _{\xi }  \notag \\
&&-(q-1)\left( q+2M-4\right) \left( \frac{2A^{\prime }}{(M+N)}\omega ^{N-1}+%
\frac{2v^{\prime }}{(M+1)}\right) \omega _{\xi }  \notag \\
&&-3(q-1)\left( \frac{A^{\prime }(N-M+2)}{(M+N)}\omega ^{N-1}+\frac{%
v^{\prime }(3-M)}{(M+1)}\right) \omega _{\xi }\quad .
\end{eqnarray}%
Comparing the last equation with (\ref{newknm}) yields the conditions for
the new coefficients 
\begin{eqnarray}
\frac{Mn}{m}A-\frac{2(q-1)\left( q+2M-4\right) }{(M+N)}A^{\prime }-\frac{%
3(q-1)(N-M+2)}{(M+N)}A^{\prime } &=&NA^{\prime } \\
\frac{M}{m}v-\frac{2(q-1)\left( q+2M-4\right) }{(M+1)}v^{\prime }-\frac{%
3(q-1)(3-M)}{(M+1)}v^{\prime } &=&v^{\prime }
\end{eqnarray}%
leading to 
\begin{eqnarray}
A^{\prime } &=&\frac{AM(M+N)}{q^{2}m(m+n)}  \label{+coeffs} \\
v^{\prime } &=&\frac{vM(M+1)}{q^{2}m(m+1)}
\end{eqnarray}%
Similar one obtains for the case (\ref{trafo-}) a $K(N,M)$ equation with
coefficients 
\begin{eqnarray}
A^{\prime } &=&\frac{vM(M+N)}{q^{2}m(m+1)}  \label{-coeffs} \\
v^{\prime } &=&\frac{AM(M+1)}{q^{2}m(m+n)}
\end{eqnarray}

One finds here that the transformation (\ref{trafo-}) interchanges the
nonlinear convection term and the dynamical term. The transformation (\ref%
{trafo-}) is thus a purely mathematical relation between the ODEs considered
with any physical implication removed.

To illustrate this result, a chart of the $K(n,m)$ equations shown in Figure %
\ref{tr-gen} gives three sets of $K(n,m)$ equations connetcted by the point
transformation discussed, i.e.\ three different equivalent classes. In fact,
under the assumption of certain kinds of traveling solutions, the latter
being specified by a particular choice of integration constants in the
potential representation, any $K(n,m)$ equation belongs to a certain
equivalence class whose representant is characterized by $m$ and $n$ with $%
(m-1)$ and $(n-1)$ having no common divisors and $n\geq 2m -1$. The latter
restriction is based on the fact that any $K(n,m)$ equation with $n<2m-1$ is
connected to a $K(n,m)$ equation with $n> 2m -1$ through a transformation (%
\ref{trafo-}).

Finally we want to mention a peculiarity arising for $m=1,\;q=2$. In this
case the first term on the r.h.s.\ in eq.\ (\ref{trafodispers}) does not
contribute and only the derivative of the potential function $\mathcal{F}%
(\omega )$ enters the transformation of the $K(n,1)$ equation. This allows
the potential function $\mathcal{F}(\omega )$ to contain an arbitrary
constant, corresponding to the term $C_{2}\omega ^{2-2m}=C_{2}$ in (\ref%
{potknm}), and according, the transformation $u=\omega ^{2}$, $\mathcal{F}%
(u) $ may additionally contain the term $4C_{2}u^{2-m}=4C_{2}u$. Here $%
\mathcal{F}(u)$ belongs to a $K(n,1)$ equation with the parameters $A$ and $v
$ and, accordingly, $\mathcal{F}(\omega )$ is the potential function of a $%
K(2n-1,1) $ equation with the parameters $A^{\prime }=n/(2(n+1))A$ and $%
v^{\prime }=v/4 $. In this way, pairs of $K(n,1)$ equations become
connected. In \cite{Rosenau-compact} this property has been adresses
especially for the pair $K(3,1)-K(5,1)$, see Fig. 4

\subsection{Example: The equivalence class of the KdV equation}

\label{kdv-ec} The KdV equation reads 
\begin{equation*}
u_{\xi \xi \xi }=A\left( u^{2}\right) _{\xi }+vu_{\xi }\quad .
\end{equation*}%
According to section \ref{potanalys}, we choose $A<0$ to have soliton
solutions. A soliton solution has the form 
\begin{equation*}
u(\xi )=\frac{3v}{2|A|}\frac{1}{\cosh ^{2}\left( \frac{\sqrt{v}}{2}\xi
\right) }\quad .
\end{equation*}%
Using the results from the precious section one can immediately state
solutions to any $K(N,1)$ and $K(N,N)$ equation. On the one hand one finds
that with the transformation $u=w^{+q}$ solutions to $K(N,1)=K(q+1,1)$
equations read (the parameters $v$ and $A$ have been expressed by the
transformed parameters $v^{\prime }$ and $A^{\prime }$) 
\begin{equation}
w(\xi )=\left( \pm \right) ^{N}\left( \frac{(N+1)v^{\prime }}{2A^{\prime }}%
\right) ^{\frac{1}{N-1}}\frac{1}{\cosh ^{\frac{2}{N-1}}\left( \frac{(N-1)%
\sqrt{v^{\prime }}}{2}\xi \right) }\quad .
\end{equation}%
As indicated by the resulting factor $(\pm )^{N}$ on the right hand side,
one finds immediately that the $K(N,1)$ equations with a symmetric potential
function, i.e.\ with odd $N$, have both soliton and anti-soliton solutions.
On the other hand the solutions for the resulting $K(N,N)=K(q+1,q+1)$
equations of the transformation $u=w^{-q}$ read 
\begin{equation}
w(\xi )=\left( \pm \right) ^{N}\left( \frac{2v^{\prime }N}{(N+1)A^{\prime }}%
\right) ^{\frac{1}{N-1}}\cosh ^{\frac{2}{N-1}}\left( \frac{\sqrt{A^{\prime }}%
(N-1)}{2N}\xi \right) \quad .  \label{nnresult}
\end{equation}%
To compare this result with the literature, e.g.\ \cite{Rosenau-knn}, one
needs to recall that the potential function belonging to the transformed $%
K(N,N)$ equation is just the negative of the usual potential function of the 
$K(n,n)$ equations used in the literature. Changing the signs accordingly
results in the change $\xi \rightarrow \mathrm{i}\xi $, i.e.\ the hyperbolic
cosine is changed to a trigonometric cosine. Eq.\ (\ref{nnresult}) then
coincides with the result of \cite{Rosenau-knn}. For $m=n=2$ the solution (%
\ref{nnresult}) can be compactified. With the appropriate changes it
represents the CSS (compare section \ref{potanalys}) 
\begin{equation}
w(\xi )=\left\{ 
\begin{array}{ll}
\displaystyle\frac{4v^{\prime }}{3A^{\prime }}\cos ^{2}\left( \frac{\sqrt{%
A^{\prime }}}{4}\xi \right) & \mathrm{for}\quad |\sqrt{A^{\prime }}\xi
/4|\leq \pi \\ 
0 & \mathrm{else}%
\end{array}%
\right. \quad .  \label{cssk22}
\end{equation}

Finally, a certain nonlinear dispersive evolution equation of non-$K(n,m)$
type will be investigated that nevertheless can be considered as an element
of the KdV equivalence class. The equation under consideration reads 
\begin{equation}  \label{comb}
\beta\left(u^2\right)_{\xi\xi\xi} + \varepsilon u_{\xi\xi\xi} = -\alpha
\left(u^2\right)_\xi - u_t\quad .
\end{equation}
With $\beta (u^2)_{\xi\xi\xi}+ \varepsilon u_{\xi\xi\xi} = \beta ((u +
\varepsilon/(2\beta))^2)_{\xi\xi\xi}$ and $u_t=-vu_\xi$, the corresponding
potential representation is found using (\ref{pfshift}). Basically we have
dealt with this problem already in section \ref{potanalys}. Thus we may
immediately assume the solution to have the form of a CSS (\ref{cssk22})
added to a constant. In particular, the potential representation reads 
\begin{equation}
\left(u_\xi\right)^2 = -\frac{\alpha}{4\beta}u^2 + \left(-\frac{\alpha
\varepsilon}{4 \beta^2}+\frac{(\alpha\varepsilon + v \beta)}{3 \beta^2}%
\right) u - C_1 - \frac{\alpha\varepsilon^2}{16\beta^3} + \frac{%
\varepsilon(\alpha\varepsilon + v\beta)}{6 \beta^3}\quad .
\end{equation}
$C_2$ is set equal to zero to avoid singularities of the potential function.
Setting 
\begin{equation*}
U(\xi) = u(\xi) + \frac{\varepsilon}{2\beta} - \frac{2\left(\alpha%
\varepsilon + v\beta \right)}{3 \alpha \beta} + \sqrt{\frac{%
4\left(\alpha\varepsilon + v\beta \right)^2}{9 \alpha^2\beta^2} - C_1 }\quad
,
\end{equation*}
the potential representation can be transformed into the form 
\begin{equation}
\left(U_\xi\right)^2 = -\frac{\alpha}{ 4\beta} U^2+ \sqrt{\frac{%
\left(\alpha\varepsilon + v\beta \right)^2}{9 \beta^4} - \frac{\alpha^2 C_1}{%
4\beta^2}}U\quad .
\end{equation}
With the preceding analysis the solution for $U(\xi)$ and thus for $u(\xi)$
can be given immediately 
\begin{eqnarray}
u(\xi) &=& \frac{4}{3\alpha\beta}\sqrt{(\alpha\varepsilon +v \beta)^2 - 9/4
\alpha^2 \beta^2 C_1} \cos^2 \left(\frac{\sqrt{\alpha/\beta}}{4}\xi\right) 
\notag \\
&& - \frac{\varepsilon}{2\beta} + \frac{2\left(\alpha\varepsilon + v\beta
\right)}{3 \alpha \beta} - \sqrt{\frac{4\left(\alpha\varepsilon + v\beta
\right)^2}{9 \alpha^2\beta^2} - C_1}\quad .
\end{eqnarray}
The solution $u(\xi)$ thus has the amplitude 
\begin{equation}
b = \frac{4}{3\alpha\beta}\sqrt{(\alpha\varepsilon +v \beta)^2 - 9/4
\alpha^2 \beta^2 C_1}
\end{equation}
and the width 
\begin{equation}
L = \frac{4}{\sqrt{\alpha/\beta}}\quad .
\end{equation}
It moves with the speed 
\begin{equation}
v^\prime = 3 \left(-\frac{\alpha \varepsilon}{4 \beta^2}+\frac{%
(\alpha\varepsilon + v \beta)}{3 \beta^2}\right)
\end{equation}
and is shifted by 
\begin{equation}
\delta = - \frac{\varepsilon}{2\beta} + \frac{2\left(\alpha\varepsilon +
v\beta \right)}{3 \alpha \beta} - \sqrt{\frac{4\left(\alpha\varepsilon +
v\beta \right)^2}{9 \alpha^2\beta^2} - C_1} \quad .
\end{equation}
For $C_1=0$ these equations simplify enomously. One finds the simple
relation for the speed 
\begin{equation}
v^\prime = \frac{3\alpha}{4 \beta}\left(\frac{\varepsilon}{\beta} -
b\right)\quad .
\end{equation}
Here one finds that $b_{crit}=\varepsilon/\beta$ constitutes a critical
amplitude. Solutions with $b>b_{crit}$ move to the left whereas solutions
with $b<b_{crit}$ move to the right. Solutions with $b=b_{crit}$ are at
rest. This property of traveling modes in systems having both quadratic and
linear dispersion is documented in \cite{Ludu-OSA,Rosenau-comb}.

\section{Summary}

In this article we have presented a potential picture for nonlinear
dispersive wave equations. The potential picture provides a simplified
representation of the original wave equation in terms of a nonlinear first
order ordinary differential equation which is valid for the set of traveling
solutions. The potential representation allows for an easy and intuitive way
to identify different types of possible solutions. It proves to be extremely
useful for the examination of nonlinear dynamical systems, since it provides
direct access to various properties of the solutions without the actual need
to solve the underlying nonlinear dispersive wave equation. A general
investigation of the potential picture reveals that solitons may exist only
in systems with linear dispersion whereas compact support solutions may
arise as possible modes in systems with quadratic dispersion. Moreover, the
specification of this concept to so-called $K(n,m)$ equations - a certain
kind of nonlinear dispersive wave equations - directly gives the relations
between the amplitude of the solitary wave and its speed or the width of the
wave and its speed and its height, respectively. Only for the compact
support solution of the $K(2,2)$ equation is the width of independent of its
speed or its height.

Furthermore, it has been shown that the potential representations of a
certain $K(n,m)$ equation can be transformed into the respective potential
representation of another $K(n,m)$-type equation by a simple point
transformation. Using the potential representation as a consistency relation
for the derivative of the wave function, $K(n,m)$ equations can be
transformed directly into another. In this way the $K(n,m)$ equations are
divided into equivalence classes each containing the set of equations that
are connected via the point transformation considered. This transformation
requires a further restriction of the admissible solutions. In addition to
the requirement of focusing on traveling solutions, the solutions are
specified by fixing the initially arbitrary integration constants to zero.
An important property of point transformations is its invertibility. All
elements of an equivalent class are uniquely connected with one another.

\section*{Acknowledgements}

This work was partially supported by the US National Science Foundation
through a regular grant (9970769) and a cooperative agreement (9720652) that
includes matching from the Louisiana Board of Regents Support Fund. U.E.
greatfully acknowledges a postdoctoral fellowship by the German Academic
Exchange Service (DAAD).

\begin{appendix}
\section{Derivation of the integrating factor}
\label{ifappend}
We start from the general nonlinear dispersive evolution equation (\ref{gen-ode}) for traveling solutions, 
\begin{equation}
\left(u^m\right)_{\xi\xi\xi} = {\cal V}(u) u_\xi\quad ,
\end{equation}
which can be integrated immediately with respect to $\xi$, giving
\begin{equation}
\left(u^m\right)_{\xi\xi} = \int_0^u {\rm d}t {\cal V}(t) + C_1
\end{equation}
where $C_1$ is an arbitrary constant. We introduce an integrating factor $f(u,u_\xi)$ that has to fulfill
\begin{eqnarray}
&&\left(u^m\right)_{\xi\xi} f(u,u_\xi) - \int_0^u {\rm d}t {\cal V}(t)f(u,u_\xi) - C_1 f(u,u_\xi)\nonumber \\
&&=\left[  \left(u^m\right)_{\xi} f(u,u_\xi) - \int_0^u {\rm d}t {\cal V}(t) F - C_1 F
\right]_\xi - \left(u^m\right)_{\xi} f_\xi(u,u_\xi) + \left[\frac{\partial}{\partial \xi}\int_0^u {\rm d}t {\cal
V}(t)\right] F\quad \quad\\ &&=\left[ a(u,u_\xi) \left(u^m\right)_{\xi} - b(u) \int_0^u
{\rm d}t {\cal V}(t)
 - C_1 F\right]_\xi
\end{eqnarray}
with $F = \int {\rm d}\xi f(u,u_\xi)$. The function $a(u,u_\xi)$ can be choosen to be $a f(u,u_\xi)$ with $a$ a
constant. Now, 
$b(u)$ and $f(u,u_\xi)$ can be determined via the differential equations
\begin{eqnarray}
a f_\xi(u,u_\xi) \left(u^m\right)_\xi &=& (1-a) f(u,u_\xi) \left(u^m\right)_{\xi\xi}\\
\int_0^u {\rm d}t {\cal V}(t) f(u,u_\xi) &=&
b^\prime (u) u_\xi \int_0^u {\rm d}t {\cal V}(t) + b(u) \left[\frac{\partial}{\partial \xi} \int_0^u {\rm d}t
{\cal V}(t)\right] u_\xi\quad . 
\end{eqnarray}
The first equation yields 
\begin{equation}
f(u,u_\xi) = \left[ \left(u^m\right)_\xi\right]^{\frac{1-a}{a}}\quad .
\end{equation}
The second equation can be easily solved with $f(u,u_\xi) = F^\prime u_\xi$ which is a consequence of  the
ansatz that $b(u)$ is a  function of $u$ only. This assumption gives $a=1/2$ and one gets for $b(u)$ and for
$f(u,u_\xi)$
\begin{eqnarray}
b(u)&=& \frac{1}{\int_0^u {\rm d}t {\cal V}(t)} \int_0^u {\rm d}t F^\prime \int_0^t {\rm d}x {\cal V}(x)\quad
,\\ f(u,u_\xi)&=& \left(u^m\right)_\xi\quad .
\end{eqnarray}

\section{Derivation of the point transformation}
\label{ptderiv}
We start from the general potential representation of the $K(n,m)$ equations, 
\begin{equation}
\left(u_\xi\right)^2 = \frac{2A}{m(m+n)} u^{n-m+2} + \frac{2v}{m(m=1)} u^{3-m} + C_1 u^{2-m} + C_2
u^{2-2m}\quad .
\end{equation}
Assuming the point transformation $u = \theta (\omega)$ one is led to
\begin{equation}
\label{trafopot}
\left(\omega_\xi\right)^2 = \frac{2A}{m(m+n)} \frac{\theta^{n-m+2}}{\theta^{\prime\,2}} + \frac{2v}{m(m+1)}
\frac{\theta^{3-m}}{\theta^{\prime\,2}} + C_1 \frac{\theta^{2-m}}{\theta^{\prime\,2}} + C_2
\frac{\theta^{2-2m}}{\theta^{\prime\,2}}\quad .
\end{equation}
Eq.\ (\ref{trafopot}) is again be the potential function of a $K(N,M)$ equation (with
different arguments $N$ and $M$), i.e.
\begin{equation}
\left(\omega_\xi\right)^2 = \frac{2A^\prime}{M(M+N)} \omega^{N-M+2} +
\frac{2v^\prime}{M(M+1)}
\omega^{3-M} + C_1^\prime \omega^{2-M} + C_2^\prime
\omega^{2-2M}\quad .
\end{equation} 
The four terms on the r.h.s.\ constitute (in combination with (\ref{trafopot})) four differential equations
that determine the transformation $\theta(\omega)$. With only two free parameters $M$ and $N$, they can not be
fulfilled simultaneously so that we have to assume $C_1=C_2=0$. One is left with
\begin{equation}
\label{T}
\begin{array}{rcl}
\displaystyle 
\frac{\theta^{n-m+2}}{\theta^{\prime\,2}} &=&\displaystyle {\cal A} \omega^{N-M+2}\\
 & &\\
\displaystyle\frac{\theta^{3-m}}{\theta^{\prime\,2}}&=&\displaystyle {\cal B} \omega^{3-M}
\end{array}
\end{equation}
with 
\begin{equation}
\begin{array}{rcl}
\displaystyle 
A^\prime &=&\displaystyle \frac{{\cal A}M(M+N)}{n(m+n)}A\\
 & &\\
v^\prime &=&\displaystyle \frac{{\cal B}M(M+1)}{m(m+1)}v
\end{array}
\end{equation}
or alternatively 
\begin{equation}
\label{T-alt}
\begin{array}{rcl}
\displaystyle \frac{\theta^{n-m+2}}{\theta^{\prime\,2}} &=&\displaystyle  {\cal B} \omega^{3-M}\\
 & &\\
\displaystyle \frac{\theta^{3-m}}{\theta^{\prime\,2}}&=&\displaystyle {\cal A} \omega^{N-M+2}
\end{array}
\end{equation}
with 
\begin{equation}
\begin{array}{rcl}
\displaystyle 
A^\prime &=&\displaystyle \frac{{\cal B}M(M+1)}{n(m+n)}A\\
 & &\\
v^\prime &=&\displaystyle \frac{{\cal A}M(M+N)}{m(m+1)}v\quad .
\end{array}
\end{equation}
In the following we will only deal with the transformations (\ref{T}). 
One finds, in general,
\begin{equation}
\label{gencond}
\theta = \left(\frac{{\cal A}}{{\cal B}} \omega^{N-1}\right)^{\frac1{n-1}}\quad .
\end{equation}
Further  we first consider the cases 
covered by $m\neq 1$, $n\neq m$, $M\neq 1$, $N\neq M$. Straightforward
integration of the equations  (\ref{T}) leads to 
\begin{eqnarray}
\theta &=& \left(\frac{1}{{\cal A}}\left(\frac{m-n}{M-N}\right)^2 \omega ^{M-N}\right)^{\frac1{m-n}}\nonumber\\
 &=& \left(\frac{1}{{\cal B}}\left(\frac{m-1}{M-1}\right)^2 \omega^{M-1}\right)^{\frac1{m-1}}\nonumber
\end{eqnarray}
These equations together with (\ref{gencond}) yield the conditions
\begin{eqnarray*}
&&\frac{N-1}{n-1} = \frac{M-n}{m-n} = \frac{M-1}{m-1}\\
&&\frac{1}{{\cal A}}\left(\frac{m-n}{M-N}\right)^2 = \frac{1}{{\cal B}}\left(\frac{m-1}{M-1}\right)^2
=1
\\
 &&\\ &&{\cal A} = {\cal B}
\end{eqnarray*}
that are fulfilled by 
\begin{eqnarray}
M-N &=& q (m-n)\\
M-1 &=& q (m-1)\\
{\cal A}&=& {\cal B} = \frac1{q^2}\quad .
\end{eqnarray}
Therefore the parameters $M$ and $N$ of the $K(N,M)$ equation belonging to the transformed
potential representation are
\begin{equation}
\label{T+}
\begin{array}{rcl}
M &=& q (m-1) +1\\
N &=& q(n-1)+1\quad .
\end{array}
\end{equation}
The transformation (\ref{T}) finally reads 
\begin{equation}
\label{T+2}
\theta = \omega^{+q}
\end{equation}
The sign of the exponent is fixed by the requirement $M,\; N>0$. 

Let us now turn to the  cases in which either $m=1$ or $m=n$ or $M=1$ or $M=N$. One finds that
the combinations
$m=M=1$ or $m-n=M-N=0$ in the case (\ref{T+}) 
are likewise covered by (\ref{T+2}). 
The remaining four combinations, i.e.\ $m=1$ and $N\neq M$, $m=1$ and  $N=M$,  $m=n$  and $N\neq M$, or $m=n$
and $M=1$, respectively, do not lead to solutions of (\ref{T}) that are in agreement with (\ref{gencond}) and
can thus be discarded. 
Equivalently one finds for the alternatively possible transformation (\ref{T-alt})
\begin{equation}
\label{T-}
\begin{array}{rcl}
M &=& q ( n-m)+1\\
N &=& q(n-1)+1\quad .
\end{array}
\end{equation}
Here the transformation reads 
\begin{equation}
\label{T-2}
\theta = \omega^{-q}\quad .
\end{equation}
In additionally to the requirement $M,\; N>0$, one has to assume in this case that $n\geq m$. 
\end{appendix}


\begin{thebibliography}{99}
\bibitem{Rosenau-Hyman} P.\ Rosenau, J.M.\ Hyman, Phys.\ Rev.\ Lett.\ 
\textbf{70} (1993) 564

\bibitem{Agrawal} G.P.\ Agrawal, Nonlinear Fiber Optics, Academic Press
(1995)

\bibitem{Ludu-cluster} A.\ Ludu, A.\ Sandulescu, W.\ Greiner, J.\ Phys.\ 
\textbf{G23} (1997) 343

\bibitem{Draayer-fission} J.P.\ Draayer, A.\ Ludu, G.\ Stoitcheva, Rev.\
Mex.\ Fis.\ 45 S2 (1999) 74

\bibitem{Rosenau-potrep} P.\ Rosenau, Phys.\ Lett.\ \textbf{A211} (1996) 265

\bibitem{Rosenau-compact} P.\ Rosenau, Phys.\ Lett.\ \textbf{A275} (2000) 193

\bibitem{Arnold} V.I.\ Arnold, Mathematical Methods of Classical Mechanics,
GTM, Springer (New York) 1978

\bibitem{Hasegawa} A.\ Hasegawa, F.\ Tappert, Appl.\ Phys.\ Lett. \textbf{23}
(1973) 171

\bibitem{Abramowitz} Handbook of Mathematical Functions, Natl.\ Bur.\
Stand.\ Appl.\ Math.\ Ser.\ No.\ 55, edited by M.\ Abramowitz and I.A.\
Stegun (U.S.\ GPO, Washington, D.C., 1964)

\bibitem{Ludu-OSA} A.\ Ludu, R.F.\ O'Connell, J.P.\ Draayer,
nlin.PS/0008026v1

\bibitem{Anderson} R.L.\ Anderson, N.Kh.\ Ibragimov, Lie-B\"acklund
transformations in applications, SIAM (Philadelphia) 1979

\bibitem{Rosenau-knn} P.\ Rosenau, Physica \textbf{D123} (1998) 525

\bibitem{Rosenau-comb} P.\ Rosenau, Phys.\ Rev.\ Lett.\ \textbf{73} (1994)
1737
\end{thebibliography}
\end{document}